      [1999/12/01 v1.4c Il Nuovo Cimento]
\documentclass[varenna]{cimento}
\usepackage{graphicx}
\title{Introduction to FFLO phases and collective mode in the
BEC-BCS crossover}
\author{R.~Combescot}
\institute{Laboratoire de Physique Statistique, Ecole Normale Sup\'erieure,
24 rue Lhomond, 75231 Paris Cedex 05, France \\
and Institut Universitaire de France}
\shortauthor{R.~Combescot}

\begin{document}

\maketitle

\section{Introduction}

My main focus in this talk will be the discussion of what might happen in a BCS superfluid when there is an imbalance between the two populations of fermionic particles forming Cooper pairs. Naturally we have in mind specifically in this conference the ultracold Fermi gases which are its subject and in particular the BEC-BCS crossover. In the last part I will consider briefly the evolution in the crossover of the collective mode arising in such a superfluid (with balanced atomic populations), which is not a completely unrelated matter.

Actually I will even much more restrain my scope since I will only tell only about what is known on this topic from past research in the domain of superconductivity, which spans almost the entire fifty years period of time since the BCS theory came out, this field being still very active with a large number of problems still open. This research is driven in particular by the search for more interesting or exotic superconductors, but also by the hope to master superconductors with extremely high critical fields, such as high T$_c$ superconductors, which are of obvious pratical interest. But it turns out to be also of interest for elementary particle physicists and astrophysicists, since similar phenomena might arise in the core of neutron stars, the involved fermions being quarks in this case. This has been reviewed recently ~\cite{ref:cana}.

This means in particular that I will not speak about the very recent huge activity in this field for ultracold gases and leave it to other speakers in this conference, such as G. Strinati and K. Levin. This is quite appropriate since I have not contributed to this activity. I have rather been interested in this problem earlier 
~\cite{ref:rc1} when it seemed to be a major problem in the way of obtaining a BCS condensation. Indeed, in contrast to the case of electrons in superconductors where there are fast spin relaxation processes leading to balanced spin populations, there are in cold gases no such processes between various hyperfine states which play the role of spins in superconductors. Hence a mixture of different hyperfine states is difficult to balance precisely, and if the critical temperature for BCS superfluidity is low, this might kill altogether the transition. Fortunately experimentalists have found a very nice way around this problem by making use of Feshbach resonance and obtaining the associated molecules. By getting rid of the remaining atoms, this ensures an exact balance between the hyperfine states populations.

Nevertheless this choice will not be so frustrating because the situation we will consider is not simple at all, and gives rise to very rich and interesting physics, so much that the only thing I will be able to do is to just give a brief overview. However it must be kept in mind that we will deal only with weakly interacting superfluids, since this is the range of validity of standard BCS theory and the physics ruling standard superconductors. This corresponds to look only, in the BEC-BCS crossover, at the BCS limit where the scattering length takes small negative values $a \rightarrow 0_{-}$. Naturally we are more interested in what may happen around unitarity, and the physics may indeed be quite different for such a strongly interacting superfluid. However there is no simple way to extrapolate toward unitarity what we will see. On the other hand this same physics can not be simply dismissed as some peculiarity of the weak coupling BCS limit, since we will see that the occuring phenomena have a fairly general origin. In particular it happens in this limit that various competing superfluid states exist only in a small range of the phase diagram. But this does not imply that the same is necessarily true around unitarity if the same kind of physics occurs. In the same spirit, it is worth recalling that states separated by a small energy difference may display very different physical properties. For example it is well known that the relative change of energy between a normal metal and its superconducting phase is typically of order 10$^{-6}$. Nevertheless the physical properties  of these two phases are vastly different.

\section{The Clogston-Chandrasekhar limit}

The natural idea which comes to mind in order to produce an imbalance between electronic populations in a superconductor is to apply a magnetic field. Indeed it will couple to the electronic magnetic moment and, by producing a difference between the chemical potentials of the two spin populations, it will induce a difference between the two spin populations. However it is well known that superconductivity disappears at a critical magnetic, but this is usually not because of the unfavorable effect of electronic population imbalance. This is due to a much stronger coupling than the one we are interested in here, namely the coupling of the magnetic field to the orbital, rather than the spin, electronic degrees of freedom. Basically the magnetic field induces supercurrents (this is the Meissner effect), and at some stage the kinetic energy associated with these currents becomes too large for the superconducting state to be energetically favorable, and the metal reverts accordingly to the normal state.

However it is also known that there are two ways for the superconductor to yield to the unfavorable effect of the magnetic field. Either it switches directly to the normal state, which is the case of type I superconductors, and this leads to a first order transition. Or, in the case of type II superconductors, the order parameter (OP) finds a more subtle way to adjust to the existence of the magnetic field, by letting some regions become normal (the OP is there equal to zero) while some others remain superconducting, with a non-zero OP. This is the famous mixed state where vortices are present and allow a partial penetration of the magnetic flux in the superconductor. Naturally for high enough magnetic field the superconducting state disappears, but this occurs more progressively by a second order transition.

This phenomenon is quite remarkable since it corresponds to a spontaneous breaking of translational invariance. What happens essentially is that the OP is flexible enough (and the OP space is large enough) to find some appropriate state to adjust to the presence of the magnetic field, rather than merely disappearing. Naturally this phenomenon is completely analogous to the appearance of vortices in an ultracold superfluid gas when it is set in rotation, as it has been observed now experimentally for fermionic as well as for bosonic gases.

The coupling of the magnetic field to the orbital degrees of freedom is due to the charge of the electron, and the resulting term in the Hamiltonian is much larger in standard superconductors than the coupling $ \mathit{E}=-{\bf M}.{\bf B}$ to the magnetic moment, which is usually completely negligible. However there are geometries where one can avoid the existence of the orbital currents, responsible for the standard critical field. Indeed if one has a quasi two-dimensional geometry, where the superconductor is essentially a stack of conducting planes, with very small tunnelling probability between planes, a magnetic field parallel to these planes would produce supercurrents flowing perpendicular to the planes. But the very small coupling between the planes will prohibit the existence of such supercurrents. This is as if the coupling to the orbital degrees of freedom had disappeared, and in this case the critical field will be controled by the above coupling of the magnetic field to the electronic magnetic moment. This kind of situation is practically extremely interesting because the resulting critical fields will be much higher than in the standard situation, and people are naturally extremely interested for applications by superconductors able to stand very high magnetic fields. It is noteworthy that the above geometry is realized in the high T$_{c}$ cuprate superconductors. Hence, apart from its fundamental interest, there is also a strong practical incentive to fully explore the physics we will consider here. However originally research in this field had been mostly driven by the quest to understand the effect of magnetic impurities in superconductors.
\begin{figure}
\vspace{-90mm}
\centerline{{\rotatebox{90}{\includegraphics[width=200mm]{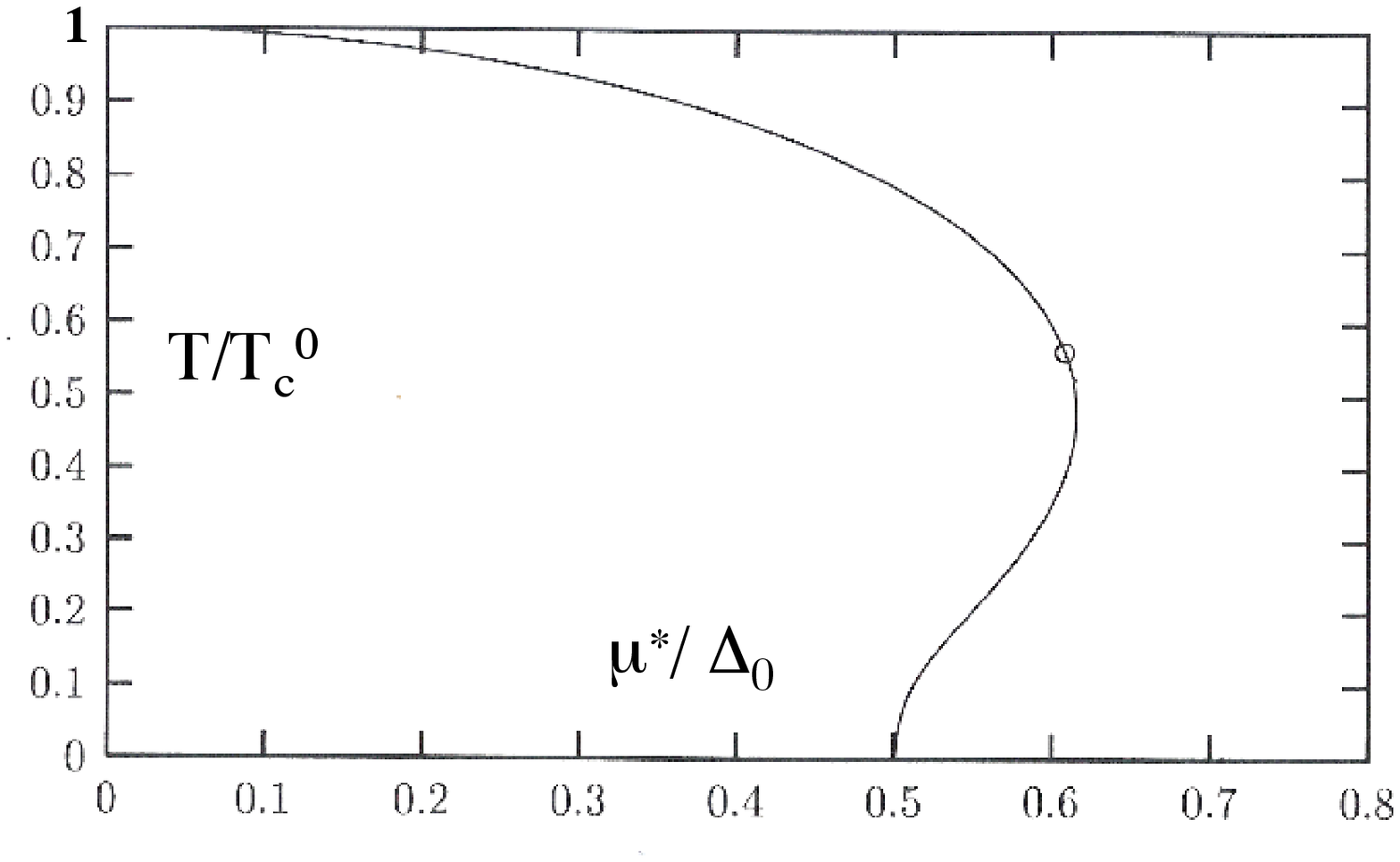}}}}
\vspace{-40mm}\caption{Temperature of the spinodal transition, compared to the critical temperature $T_{c}^{0}$ for equal populations, as a function of the chemical potential difference $\mu ^{*}$ expressed in terms of the zero temperature gap $\Delta _{0}$ for equal populations}
\label{fig1}
\end{figure}

It is clear that an imbalance $n_{\uparrow} \neq n_{\downarrow}$ between the two spin populations, as created by a difference between spin up chemical potential $\mu _{\uparrow}$ and spin down chemical potential $\mu _{\downarrow}$ is unfavorable to the BCS state since the formation of Cooper pairs imply naturally $n_{\uparrow} = n_{\downarrow}$. In the following we will consider the difference $\mu _{\uparrow} - \mu _{\downarrow} \equiv 2 \mu ^{*}$ as our primary external constraint on the system, because it is easier to handle. It is then a matter of standard thermodynamic translation in terms of conjugate variable to obtain the effect of an imbalance in the spin populations, which is the natural variable ofr cold gases. Qualitatively we can think of $\mu ^{*}$ as an effective magnetic field and its unfavorable effect on pairing implies that, beyond some critical chemical potential difference $\mu ^{*}_{c}$, the BCS state will be destroyed in favor of the normal state. In order to find the dependence on temperature $\mu ^{*}_{c}(T)$ of this critical "field", it is natural to generalize the standard BCS equation which gives the critical temperature to the case where $\mu ^{*} \neq 0$. One finds easily:
\begin{eqnarray}\label{eqga}
\frac{1}{N_{0}V}=  \int_{0}^{\omega _c} \frac{1-f(E_{k,\uparrow})-f(E_{k,\downarrow})}{2E_k}
\end{eqnarray}
where $f(E)=[\exp(E/T) -1]^{-1}$ is the Fermi distribution, $\omega _c$ the standard cut-off of BCS theory, $N_{0}$ the single spin density of states at the Fermi surface and $V$ the attractive interaction, responsible for pairing. As usual the standard single particle excitation energy $E_k = \sqrt{\xi _{k}^{2}+\Delta ^{2}}$ is related to the kinetic energy $\epsilon _k = \hbar^{2}k^{2}/2m$ and $\xi_k=\epsilon _k -\mu$ is just this kinetic energy measured from the averaged chemical potential $\mu = (\mu _{\uparrow} + \mu _{\downarrow})/2$, while the gap $\Delta $ is zero at the critical field we are looking for. The effect of $\mu ^{*}$ enters this equation only because the energies entering the Fermi distributions are the up spin and down spin single particle excitation energies $ E_{k,\uparrow,\downarrow}=\pm \mu ^{*} +E_k$.

It is worth pointing out that the transition corresponding to this equation is just a natural continuation of the standard second order phase transition from the normal phase toward the BCS phase at $\mu ^{*} =0$. It corresponds to an absolute instability of the normal phase with respect to the superconducting one, and is often called a spinodal instability. There is no supercooling effect at this transition, and the normal state can not possibly escape to go to the superfluid one.
The result is shown on fig.\ref{fig1}. At $T=0$ one finds the simple result $\mu ^{*}=\Delta_{0}/2$, where $ \Delta_{0}$ is the gap at zero temperature when the spin populations are equal. However the result displays a rather strange reentrant behaviour, with the critical field at intermediate temperature being larger than the one obtained at $T=0$.

However it was realized quite early by Clogston and Chandrasekhar~\cite{ref:clog,ref:chand}, independently, that there is at $T=0$ a first order transition from the normal to the standard BCS superfluid state occuring at a higher field. The location of this transition is easy to obtain. At zero temperature, for fixed chemical potentials  $\mu _{\uparrow} $ and $\mu _{\downarrow}$, the stable phase is given by the minimum of the thermodynamic potential:
\begin{eqnarray}\label{}\hspace{-60mm}
G = E - \mu _{\uparrow}n_{\uparrow} - \mu _{\downarrow}n_{\downarrow }
\end{eqnarray}
\begin{eqnarray}\hspace{20mm}
= E  - \frac{\mu _{\uparrow} + \mu _{\downarrow}}{2}\; (n_{\uparrow}+n_{\downarrow}) -  \frac{\mu _{\uparrow} - \mu _{\downarrow}}{2} \;(n_{\uparrow}-n_{\downarrow})
= E - \mu ^{*}\,(n_{\uparrow}-n_{\downarrow}) \nonumber
\end{eqnarray}
where in the last equality we have just used the definition of $\mu ^{*}$ and dropped for clarity the term corresponding to the total spin population since it does not play here any role. Now we have in the standard BCS state $n_{\uparrow}=n_{\downarrow}$, while its energy $E$ is given in terms of the zero temperature gap $\Delta _{0}$ by the standard result $E=-(1/2)N_{0} \Delta_{0}^{2}$ compared to the normal state energy, which leads to
\begin{eqnarray}
G_{s} = - \frac{1}{2}\,N_{0} \Delta_{0}^{2}
\end{eqnarray}
On the other hand we have taken by definition the energy of the normal state equal to zero for equal populations and, in contrast with the BCS state, the normal state has a non zero susceptibity leading to $n_{\uparrow}-n_{\downarrow}= 2 N_{0}\,\mu ^{*}$. As a result we have $E= N_{0} \mu^{*2}$ and:
\begin{eqnarray}\label{}
G_{n} = - N_{0} \, \mu ^{*2}
\end{eqnarray}
Clearly, for $\mu ^{*}=0$, the BCS state has the lower $G$ while for large $\mu ^{*}$ this will be the normal state, as expected. The transition occurs for $G_{n} = G_{s}$, which leads for the critical $\mu ^{*}$ to
\begin{eqnarray}\label{clch}
\mu ^{*}_{c} = \frac{\Delta_0 }{\sqrt{2}}
\end{eqnarray}
Here the transition is first order since the gap, which is the OP in the present case, goes discontinously from zero to $\Delta_0$, when one goes from the normal to the superfluid state. Since the above $\mu ^{*}_{c} =\Delta_0 / \sqrt{2}$ is higher than the one, equal to $\Delta_0 /2$, that we have found from eq.\ref{eqga}, it is clear that, coming from the normal state with a high $\mu ^{*}$, the transition at  $\mu ^{*}_{c} =\Delta_0 / \sqrt{2}$ will occur first. However supercooling below this transition is possible, since it is first order. This may occur down to $\Delta_0 /2$ where the normal state becomes absolutely unstable with respect to the BCS state.
\begin{figure}
\vspace{-30mm}
\centerline{{\rotatebox{270}{\includegraphics[width=200mm]{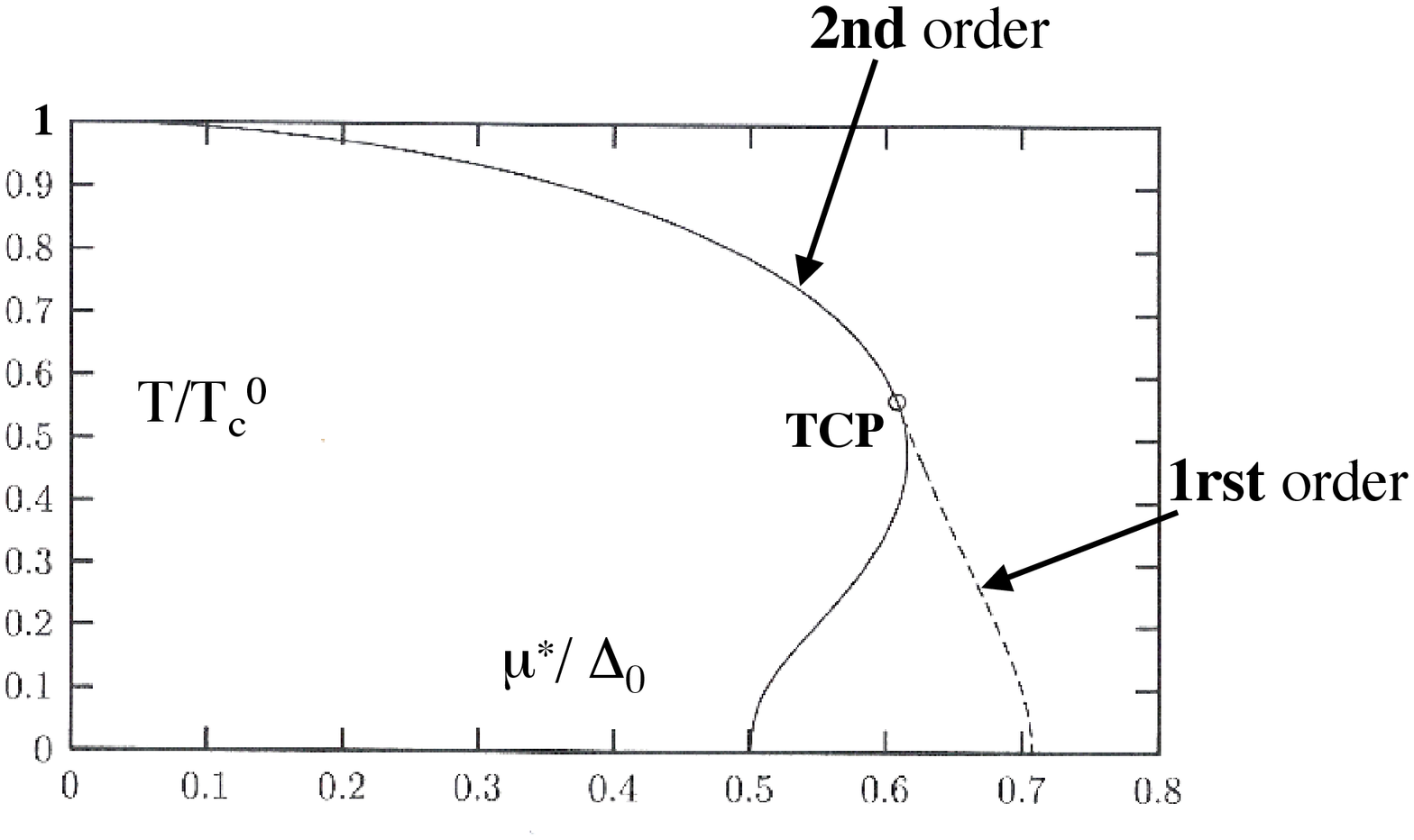}}}}
\vspace{-100mm}
\caption{Temperature of the phase transition between normal and BCS state. The dashed line gives the location of the first order transition corresponding to the Clogston-Chandrasekhar limit. The full line is the same as in fig.1. Above the tricritical point (TCP) it gives the transition which is second order. Below the TCP, it corresponds to the position of the spinodal line where the normal state is completely unstable with respect to the transition toward the BCS state.}
\label{fig2}
\end{figure}

The resulting transition line is given now in fig.\ref{fig2}. When the $T=0$ first order transition we has just found is followed at non zero temperature, one finds that this line meets the second order transition line resulting from eq.\ref{eqga} for some intermediate temperature at the so-called tricritical point (TCP). Hence, at this stage, we have reached the conclusion that the transition from normal to superfluid is second order above the TCP, but becomes first order at lower temperatures. As we will see the actual situations is much more complicated, but this TCP plays an important role in our understanding, and we will consider it again below.
\begin{figure}
\vspace{-40mm}
\centerline{{\rotatebox{270}{\includegraphics[width=200mm]{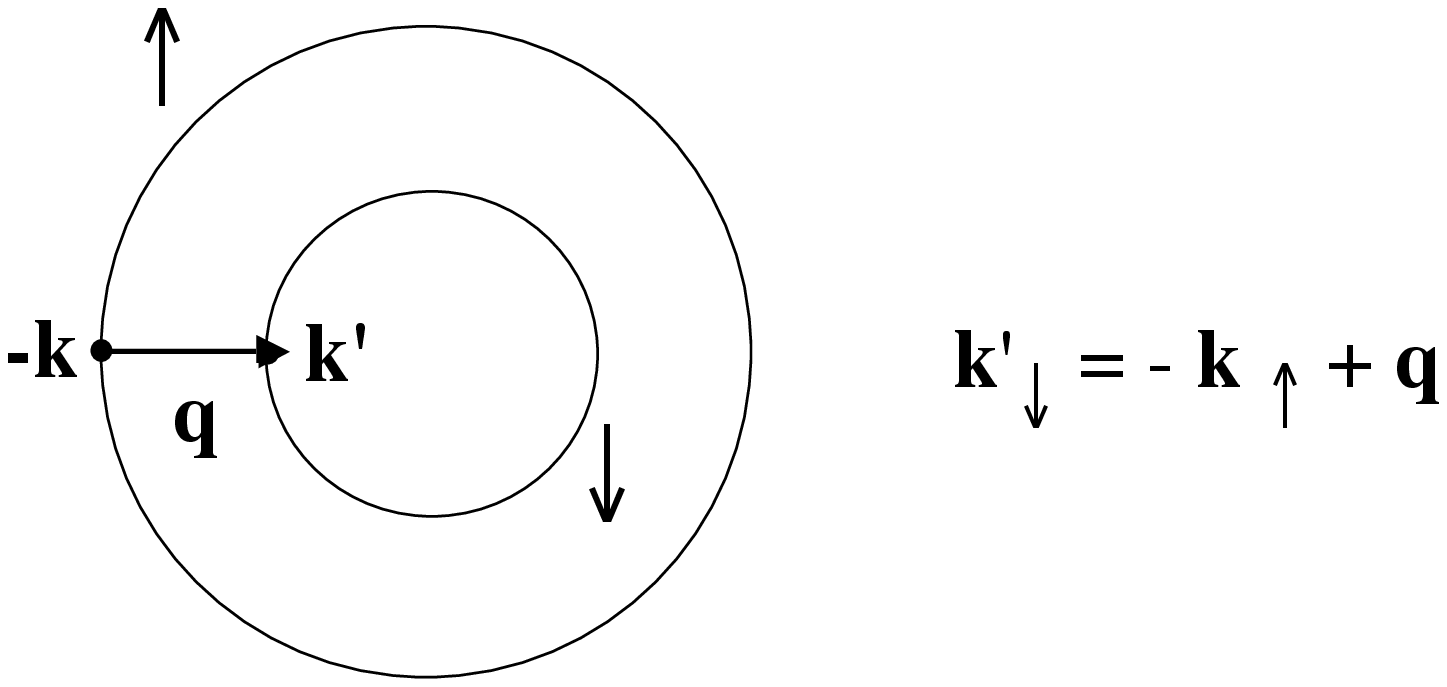}}}}
\vspace{-110mm}
\caption{Fermi surfaces of up and down spins when their chemical potential are different}
\label{fig3}
\end{figure}
\section{The Fulde - Ferrell - Larkin - Ovchinnikov phases}

Not much later the independent work of Fulde and Ferrell~\cite{ref:ff}, and of Larkin and Ovchinnikov~\cite{ref:lo} showed that the actual transition is more complex. Indeed they considered the possibility for the Cooper pairs to have a non zero total momentum ${\bf q}$. This is in contrast with the standard BCS phase where it is energetically more favourable to have all the pairs with a total momentum ${\bf q}=0$. They found that, at $T=0$, for high effective field $\mu ^{*}$, the transition toward the superfluid state toward an order parameter with ${\bf q} \neq 0$, is energetically more advantageous, leading in this way to an extension of the superfluid stability domain in the presence of an effective field $\mu ^{*}$. 

This result is quite remarkable since it corresponds, as we will see below in more details, to a spontaneous breaking of translational invariance produced by the field $\mu ^{*}$. This is quite similar to what occurs in a type II superconductor in the presence of a magnetic field $B$, as we have seen above. The superfluid phases having such an order parameter with ${\bf q} \neq 0$ are called FFLO or LOFF phases. However, despite many experimental efforts and a good deal of claims, these phases have not yet been observed in standard superconductors, at least very clearly. Nevertheless the existence of such an effect seems quite real. This is perhaps most convincingly shown by experiments where a nearby magnetic material was responsible for the effective field $\mu ^{*}$. This was felt by a superconductor, due to the proximity effect. The existence of a resulting change of sign of the OP could be proved by the study of the characteristics of the associated tunnel junction~\cite{ref:kontos}.
\begin{figure}
\vspace{-40mm}
\centerline{{\rotatebox{270}{\includegraphics[width=200mm]{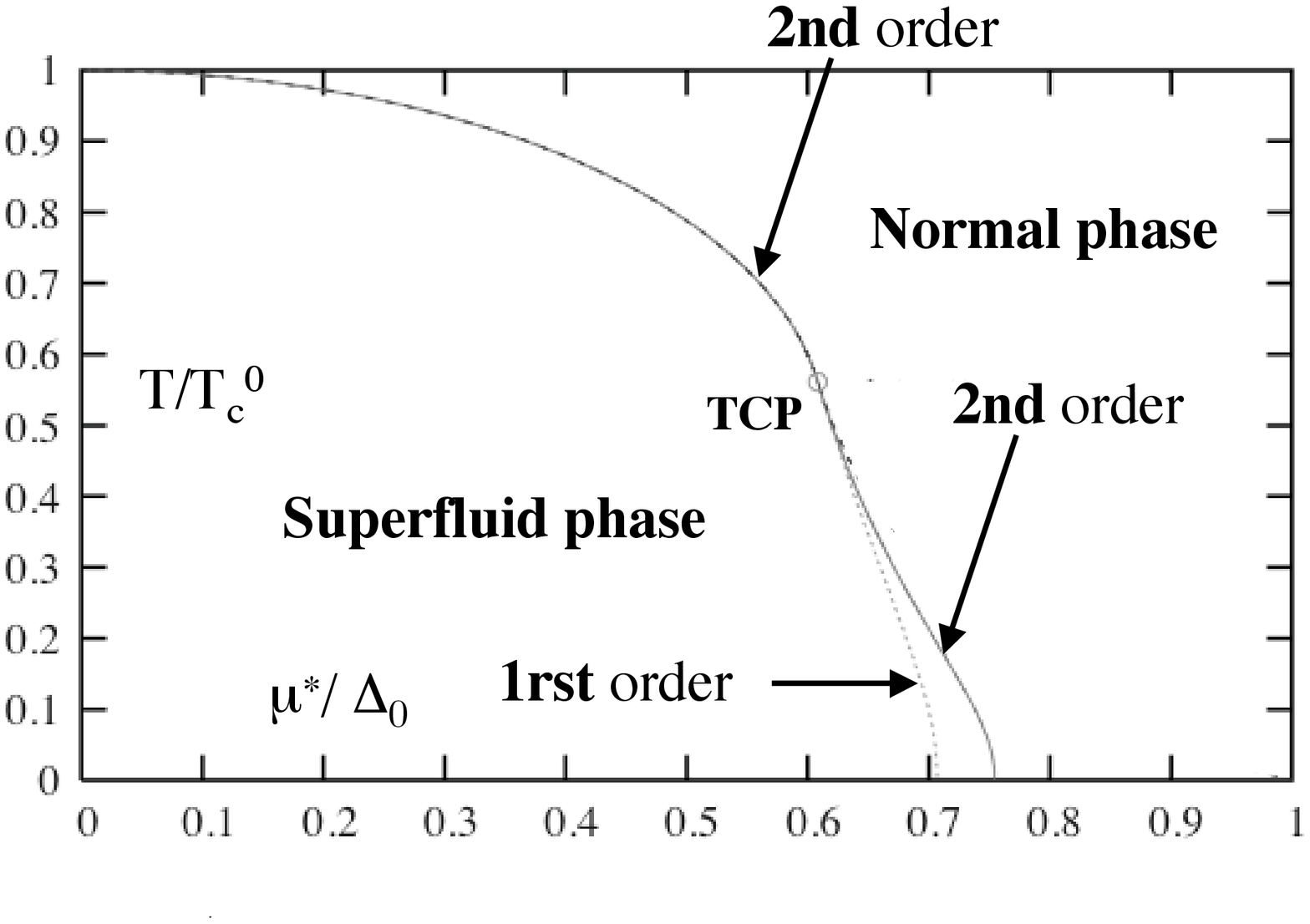}}}}
\vspace{-80mm}
\caption{Temperature of the phase transition between normal and BCS state. The part above the TCP is the same as in fig.2, and below the TCP the dashed line is again the location of the first order Clogston-Chandrasekhar limit. The full line below the TCP is the FFLO second order phase transition.}
\label{fig4}
\end{figure}

Qualitatively one can understand in the following way the energetical advantage of having Cooper pairs with a total momentum ${\bf q} \neq 0$. There is always a loss in kinetic energy for electrons in forming Cooper pairs, compared to the kinetic energy of a simple $T=0$ Fermi sea, because one has to excite electrons away from the Fermi surface in order to build up the Cooper pair wavefunction. Naturally this loss is more than compensated by the gain in attractive interaction energy obtained by this formation. However in order to minimize this kinetic energy loss, it is better to have the electrons as much as possible near the Fermi surface. When the up and down spin Fermi surfaces are identical this requirement is completely compatible with the standard formation of ${\bf k}_{\uparrow},{\bf -k}_{\downarrow}$ pairs. However when the up and down spin chemical potentials are different, the Fermi surfaces are also different and it is no longer possible to take each electron in the vicinity of its Fermi surface while having at the same time a zero total momentum. As shown in fig.\ref{fig3}, taking each electron in the vicinity of its Fermi surface implies a non zero total momentum $ {\bf k}_{\uparrow} + {\bf k}'_{\downarrow}= {\bf q}$. Hence taking Cooper pairs with non zero total momentum allows to better cope with the requirement of minimizing the kinetic energy.

However, although this argument shows why pairs with ${\bf q} \neq 0$ are coming in, it does not prove that this is energetically favourable globally. Indeed, since total momentum is conserved by the interaction, we have to choose a specific total momentum ${\bf q}$, independent of ${\bf k}$. But, if as shown on fig.\ref{fig3}, a given choice of ${\bf q}$ allows to take both electrons in the vicinity of their respective Fermi surface for some wavevectors ${\bf k}$, this is not true for all wavevectors (since one Fermi surface is not related to the other by a simple translation ${\bf q}$): on fig.\ref{fig3} this would not work at all on the opposite sides of the Fermi surfaces. Hence a quantitative calculation is necessary to obtain the final answer.

The first step was done by Fulde and Ferrell~\cite{ref:ff}, and Larkin and Ovchinnikov~\cite{ref:lo}. Looking for a second order phase transition, they found indeed that taking ${\bf q} \neq 0$ is globally favorable. Specifically they obtained that at $T=0$ the transition occurs for a critical field $\mu ^{*}=0.754 \,\Delta _{0}$ with a corresponding wavevector $q=2.40\, \mu ^{*}/v_F$, where $v_F$ is the Fermi velocity (in the weak coupling regime, the two Fermi surfaces are only slightly different and, to lowest order, they have the same $v_F$). This value of $q$ implies that the order parameter has spatial variations with wavelength of order of the size of a Cooper pair. Nevertheless we see that the resulting critical field is only barely above the Clogston-Chandrasekhar result. In this respect it is worth noting that this situation is much improved when one goes to lower spatial dimensions, which is relevant for the planar structure of high T$_{c}$ superconductors and could also be realized with cold gases.

Although the modulus of ${\bf q}$ is fixed by the energy minimization, its direction is not and we have naturally a degeneracy with respect to this direction, linked to the spontaneous breaking of rotational invariance, but also with respect to any superposition of similar plane waves. The work of Fulde and Ferrell~\cite{ref:ff} concentrated on the case where a single wavevector ${\bf q}$ is chosen. This corresponds to an order parameter $\Delta ({\bf r})$ which has a single plane-wave form $\Delta ({\bf r}) \sim \exp(i{\bf q}.{\bf r})$. Larkin and Ovchinnikov~\cite{ref:lo} went further by noticing that, just below the transition, in a standard Ginzburg-Landau analysis, the fourth order term would produce couplings between different plane waves which would lift the degeneracy and select a specific order parameter.

In order to be slightly more specific, let us first recall the standard expression of the free energy $F$ in terms of an expansion in powers of the order parameter $\Delta $, assumed first to be space independent:
\begin{eqnarray} \label{fr1}
F = a_0 \Delta ^{2} + a_2 \Delta ^{4}
\end{eqnarray}
where the normal state free energy is taken to be zero. The coefficient $a_2$ is always positive, while we have $a_0 > 0$ above and $a_0 < 0$ below the critical temperature. In a $T=0$ analysis, the free energy is merely the energy, and the role of the temperature is played by the effective field $\mu ^{*}$.
Now, because of the degeneracy with respect to the momentum direction, the order parameter can be any superposition of plane waves:
\begin{eqnarray}\label{pw}
\Delta ({\bf r}) =  \sum_{{\bf q}} \,\Delta _{{\bf q}}\;e^{i{\bf q}.{\bf r}}
\end{eqnarray}
provided the modulus $q$ is fixed to the above mentionned value. The generalization to this case of the above Ginzburg-Landau expansion for the free energy is:
\begin{eqnarray}
F = \sum_{{\bf q}} a_0 |\Delta _{{\bf q}}| ^{2} + \sum_{{\bf q}_1 +{\bf q}_3 = {\bf q}_2 +{\bf q}_4 } J({\bf q}_1,{\bf q}_2,{\bf q}_3,{\bf q}_4) \Delta_{{\bf q}_1} \Delta ^{*}_{{\bf q}_2}\Delta_{{\bf q}_3} \Delta ^{*}_{{\bf q}_4}
\end{eqnarray}
where $a_0$ depends only on the modulus of ${\bf q}$ by rotational invariance, but we have not written it explicitely since this modulus is fixed. On the other hand $J({\bf q}_1,{\bf q}_2,{\bf q}_3,{\bf q}_4)$ depends on the relative orientation of the various wavevectors and minimization will produce a selection among the possible orientations. Indeed $a_0$ as well as $J({\bf q}_1,{\bf q}_2,{\bf q}_3,{\bf q}_4)$ can be explicitely calculated within BCS theory. But we see that this problem is immediately extremely complicated since we want to minimize the free energy with respect to all the possible $\Delta _{{\bf q}}$, which means that we have a functional space to explore.

Larkin and Ovchinnikov did simplify it somewhat by the following reasonable assumption. The superposition of the various plane waves will give rise, by constructive interference, to maxima of $\Delta ({\bf r})$(correspondingly at these points the difference in spin populations will be minimal). Larkin and Ovchinnikov assumed that these would be regularly arranged, as in a cristalline lattice. They selected the subset of wavevectors ${\bf q}_i$ giving rise to such a structure. With this restriction they could perform a complete exploration and came to the conclusion that the best structure was obtained by taking just the equal superposition of two plane waves with opposite wavevectors. In other words the best order parameter is given by:
\begin{eqnarray}
\Delta ({\bf r}) \sim\;\cos ({\bf q}.{\bf r})
\end{eqnarray}
where naturally a degeneracy with respect to global rotation is left, but no longer with respect to any superposition of plane waves. Let us stress again that this Ginzburg-Landau analysis assumes the transition to be second order.

The transition line is easy to follow at non zero temperature because all possible FFLO phases have the same critical temperature. When this is done one finds, as shown on fig.\ref{fig4}, that this line follows rather closely the Clogston-Chandrasekhar line and meets it at the tricritical point, which we have already considered above. This makes it clearly worthwhile to consider more closely the vicinity of this tricritical point, which will gives us a much better understanding of how the various transitions we have already discussed are related. I will sketch in the next section the results of the analytical study which has been done with C. Mora~\cite{ref:rcmora}.

\section{Vicinity of the tricritical point}
\begin{figure}
\vspace{-40mm}
\centerline{{\rotatebox{270}{\includegraphics[width=200mm]{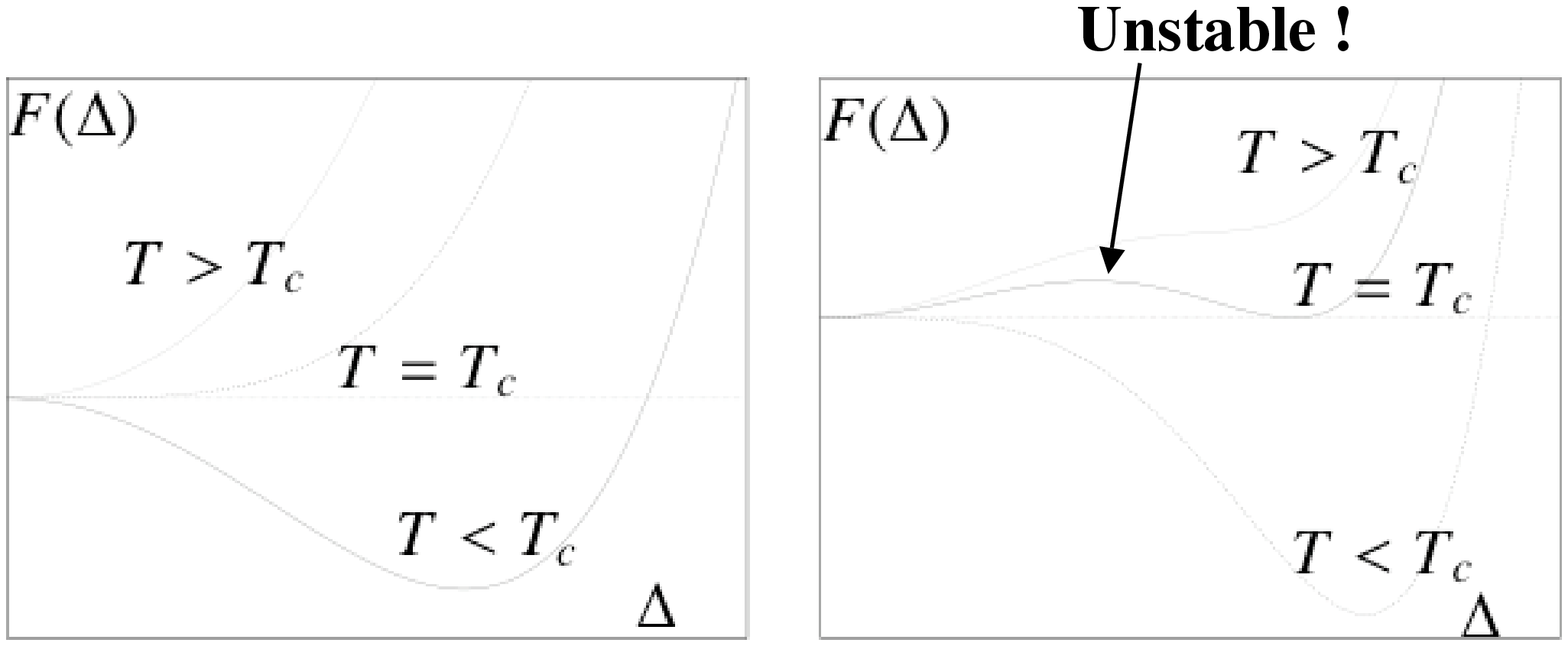}}}}
\vspace{-90mm}
\caption{Free energy $F(\Delta )$ as a function of $\Delta $ given by eq.\ref{fr2} in the cases where $a_2$ is positive (left panel) or negative (right panel), for temperatures above, below or at the corresponding critical temperature $T_c$.}
\label{fig5}
\end{figure}
This domain is particularly interesting because we will be able to make a generalized Ginzburg-Landau analysis for the following reasons. First since we know that, at temperatures above the TCP, the transition is just the standard second order BCS transition, we know by continuity that the order parameter will be small anyway in the vicinity of the TCP. This is true not only naturally if the transition is second order, but also if it is first order. This point is particularly interesting because we will be able to study the possibility of a first order transition, whereas this is in general quite difficult since the order parameter is not small by definition and no expansion in its powers are accordingly possible. The second reason has to do with the spatial dependence of the order parameter. We know that, at temperatures above the TCP, we have the standard uniform BCS phase. Again by continuity we know that the spatial variation of the order parameter (which occurs in the FFLO phases) will have only long wavelength and small wavevector, which will allow us to perform an expansion in powers of this wavevector ${\bf q}$. It turns out that an expansion up to sixth order in powers of the couple of variable $(\Delta ,{\bf  q})$ is enough.

In order to clarify the fairly complex situation we want to consider, let us first display the variation of the free energy for an homogeneous order parameter, when one has to go up to the sixth order term in the expansion which is not the standard Ginzburg-Landau expansion we have seen above. In this case the expression of the free energy is:
\begin{eqnarray}\label{fr2}
F = a_0 \Delta ^{2} + a_2 \Delta ^{4} + a_4 \Delta ^{6}
\end{eqnarray}
where again all the coefficients can be calculated explicitely within BCS theory. In this expression $a_4$ is always positive in the vicinity of the TCP, which is the basic reason why we can stop the expansion at order six. As above in eq.\ref{fr1} the coefficient $a_0$ is positive above the second order transition and negative below it. The new point is the behaviour of $a_2$. Above the TCP this coefficient is positive as was the case in eq.\ref{fr1}. However it decreases when we come near the TCP and it changes sign when we cross the TCP. This is actually the essential property which gives rise to the TCP: at this point we have $a_2=0$.

We show in fig.\ref{fig5} $F(\Delta)$ as given by eq.\ref{fr2}. On the left panel we see the case where $a_2>0$, for decreasing values of $a_0$. Qualitatively there is no difference with what one obtains from eq.\ref{fr1}, because the $a_4$ term just reinforces the effect of the $a_2$ term, and in particular the transition occurs for $a_0=0$ and is second order. On the other hand we see on the right panel that the situation is qualitatively modified when $a_2<0$. Indeed this introduces a downward bending of the graph and as a result, when $a_0$ is lowered, a minimum develops for a non zero value of $\Delta $. When the value of $F(\Delta)$ reaches zero at this minimum, due to the lowering of $a_0$, there is a transition toward the superfluid phase, which is a first order transition since the value of $\Delta $ will be non zero. Accordingly we see that, because of the change of sign of $a_2$, the standard second order phase transition is superseded by a first order transition.

At this stage it is worth making a side remark on the curve for $T=T_c$ on the right panel. It has two minima, one for $\Delta =0$ which corresponds to the normal state, and another one at non zero $\Delta $ which corresponds to the superfluid state. Naturally we find in between these two minima a maximum of the free energy. Now the gap equation of BCS theory can be obtained quite generally by writing that the free energy is extremal with respect to the order parameter $\Delta $. This means that, in the present situation, this gap equation will have three solutions corresponding to the two minima and to the maximum. In the course of his study of the generalization at non zero temperature of the Clogston-Chandrasekhar limit, Sarma~\cite{ref:sarma} found indeed three solutions (we will see below that the Clogston-Chandrasekhar limit fits indeed in our framework). He studied in particular the one corresponding to the maximum, which is often referred to as the Sarma phase in the recent literature.
However this phase does not appear in the final thermodynamic results of Sarma's paper. This is expected since it is not a physically acceptable solution because it corresponds to a maximum, and not a minimum of the free energy. This is naturally the situation in the weak coupling BCS limit, and the situation might be different in cold gases away from weak coupling. However it is clearly necessary to check in these cases that the Sarma phase found in the course of calculations corresponds to a minimum, and not a maximum, of the free energy.

Let us come back now to our problem and take into account that, having mind FFLO phases, we want to consider space dependent order parameters which we can always Fourier expand as in eq.\ref{pw}. We restrict ourselves to the interesting domain just below the TCP. Since the resulting expression for the free energy is fairly complicated, we write it with reduced units (but we do not change the notations for simplicity). Without entering in the details, we set $a_0 \sim (a^{2}_2/ a_4) A_0$ , and $q^2$ as well as $\Delta^2$ are also expressed in units proportional to $a_2 / a_4$. The reduced coefficient $A_0(\mu ^{*},T)$, obtained from BCS theory, is an increasing function of effective field $\mu ^{*}$ and temperature $T$ in the domain we are interested in. This reduced free energy reads (the normal state free energy being taken as zero): 
\begin{eqnarray}\label{fr3}
\hspace{-70mm}F = \sum_{{\bf q}} |\Delta _{{\bf q}}| ^{2} \left[A_0 - \frac{1}{3} q^2 + \frac{1}{5} q^4\right]  \nonumber
\end{eqnarray}
\vspace{-5mm}
\begin{eqnarray}
- \frac{1}{4} \sum  \Delta_{{\bf q}_1} \Delta ^{*}_{{\bf q}_2}\Delta_{{\bf q}_3} \Delta ^{*}_{{\bf q}_4}
\left[1 - \frac{1}{3}(q_{1}^{2}+q_{2}^{2}+q_{3}^{2}+q_{4}^{2}+{\bf q}_1.{\bf q}_3 + {\bf q}_2.{\bf q}_4 )\right]
 \nonumber
\end{eqnarray}
\vspace{-3mm}
\begin{eqnarray}
+ \frac{1}{8} \sum  \Delta_{{\bf q}_1} \Delta ^{*}_{{\bf q}_2}\Delta_{{\bf q}_3} \Delta ^{*}_{{\bf q}_4}\Delta_{{\bf q}_5} \Delta ^{*}_{{\bf q}_6}
\end{eqnarray}
In this expression are present all the transitions we have discussed precedingly. Indeed the
standard BCS spinodal transition corresponds to a uniform order parameter, so we have ${\bf q}=0$, and is a second order transition, that is $\Delta =0$. Accordingly its location in the ($\mu ^{*},T$) plane is given by $A_0 = 0$. However we know that a first order transition to a uniform order parameter with $\Delta  \neq 0$, as found by Clogston-Chandrasekhar, is more favorable. And indeed we find that, for ${\bf q}=0$, when $A_0 = 1/8$, the minimum found at $\Delta = 1$ has already a free energy $F=0$ (this corresponds to the minimum of $F(\Delta )$ at $T_c$ on the right panel of fig.\ref{fig5}). The fact that the corresponding value $1/8$ of $A_0$ is larger than zero implies that this transition occurs (for fixed $\mu ^{*}$) at higher temperature than the standard BCS transition, so that it supersedes it.

Now we can consider what occurs when we take into account the possibility of an order parameter with ${\bf q} \neq 0$. We look first for a second order transition, that is $\Delta =0$. We see immediately from the first term in eq.\ref{fr3}, of order $\Delta ^2$, that it is indeed better to take $q^2= 5/6$ which gives the minimum of the $q$-dependent coefficient. This leads to $A_0 = 5/36$ which is even higher than the Clogston-Chandrasekhar value $A_0 = 1/8$. Therefore this FFLO transition overtakes the Clogston-Chandrasekhar one. It is interesting to note, at this stage, that there is no barrier in free energy between $q=0$ and $q^2= 5/6$, as it is obvious from the first term of eq.\ref{fr3}. This implies that the transition to a $q \neq 0$ order parameter has a spinodal character. It is impossible to escape it because it corresponds to an absolute instability. In this weak coupling limit it is not possible to believe that the FFLO phases will be "missed" by the system. This is directly linked to the fact that the coefficient of the $q^2$ term is negative. If we do  not take reduced units, this coefficient is actually found proportional to $a_2$. Above the TCP it will be positive and we are back to the standard BCS transition to a uniform order parameter. On the other hand below the TCP it is negative and we find FFLO phases. In other words a characteristic property of the TCP is also the fact that this coefficient of the $q^2$ is equal to zero at the TCP.
\begin{figure}
\vspace{-30mm}
\centerline{{\rotatebox{270}{\includegraphics[width=200mm]{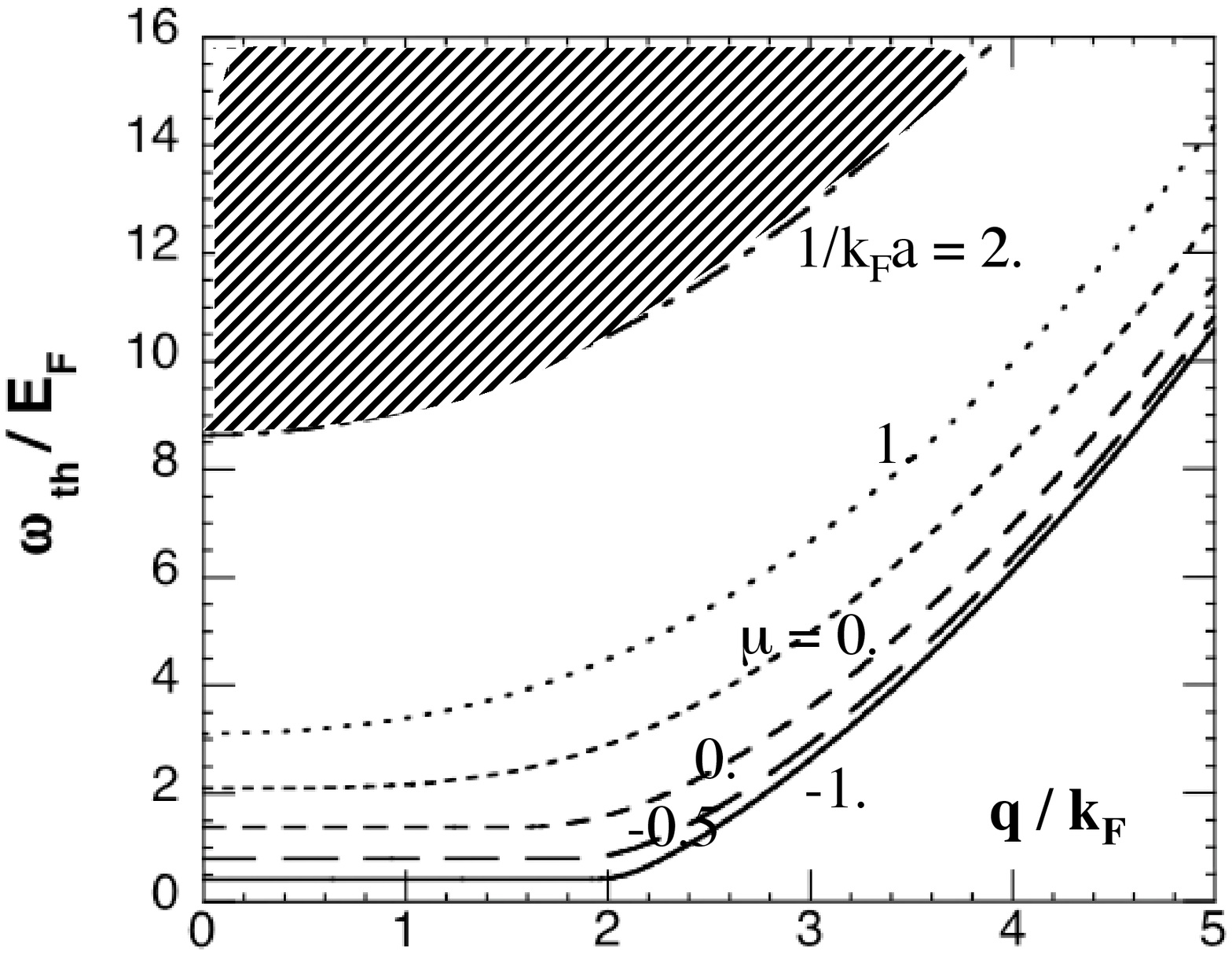}}}}
\vspace{-70mm}
\caption{Threshold $\omega _{th}$ as a function of wavevector $q$ for, from bottom to top, $1/k_Fa=$ -1. , -0.5 , 0. (unitarity) , 0.553 ($\mu =0$) , 1. and 2.}
\label{fig6}
\end{figure}

Finally when we put together all the results found above, we may naturally wonder if it is not possible to find an even better solution by looking for a first order transition $\Delta \neq 0$ at $q \neq 0$. This is indeed what we have found~\cite{ref:rcmora}. For $q^2= 0.68$ and $\Delta  = 0.27$, one finds a minimum equal to zero in the free energy if one takes $A_0 = 5/36 + 2. 10^{-3}$, which is slightly higher than the FFLO value. Hence this first order transition supersedes the standard FFLO second order phase transition. However in contrast with this standard FFLO transition, this result is not obtained for any superposition of plane waves. We have in eq.\ref{fr3} to take advantage of the specific dependence of the fourth order term on the directions of the various ${\bf q}_i$, in much the same way as in the Larkin-Ovchinnikov analysis. We find that the above result is obtained for an order parameter which has the same spatial dependence, namely $\Delta ({\bf r}) \sim\;\cos ({\bf q}.{\bf r})$, as the one obtained by Larkin and Ovchinnikov. The fact that the result for $A_0$ is barely higher than the standard FFLO one  is somewhat puzzling (which implies that the transition lines are very close; in fig.\ref{fig4} this first order transition line is just above the FFLO one). However we note that the value of $\Delta $ itself is not particularly small, compared to the Clogston-Chandrasekhar value.
\begin{figure}
\vspace{-90mm}
\centerline{{\rotatebox{90}{\includegraphics[width=200mm]{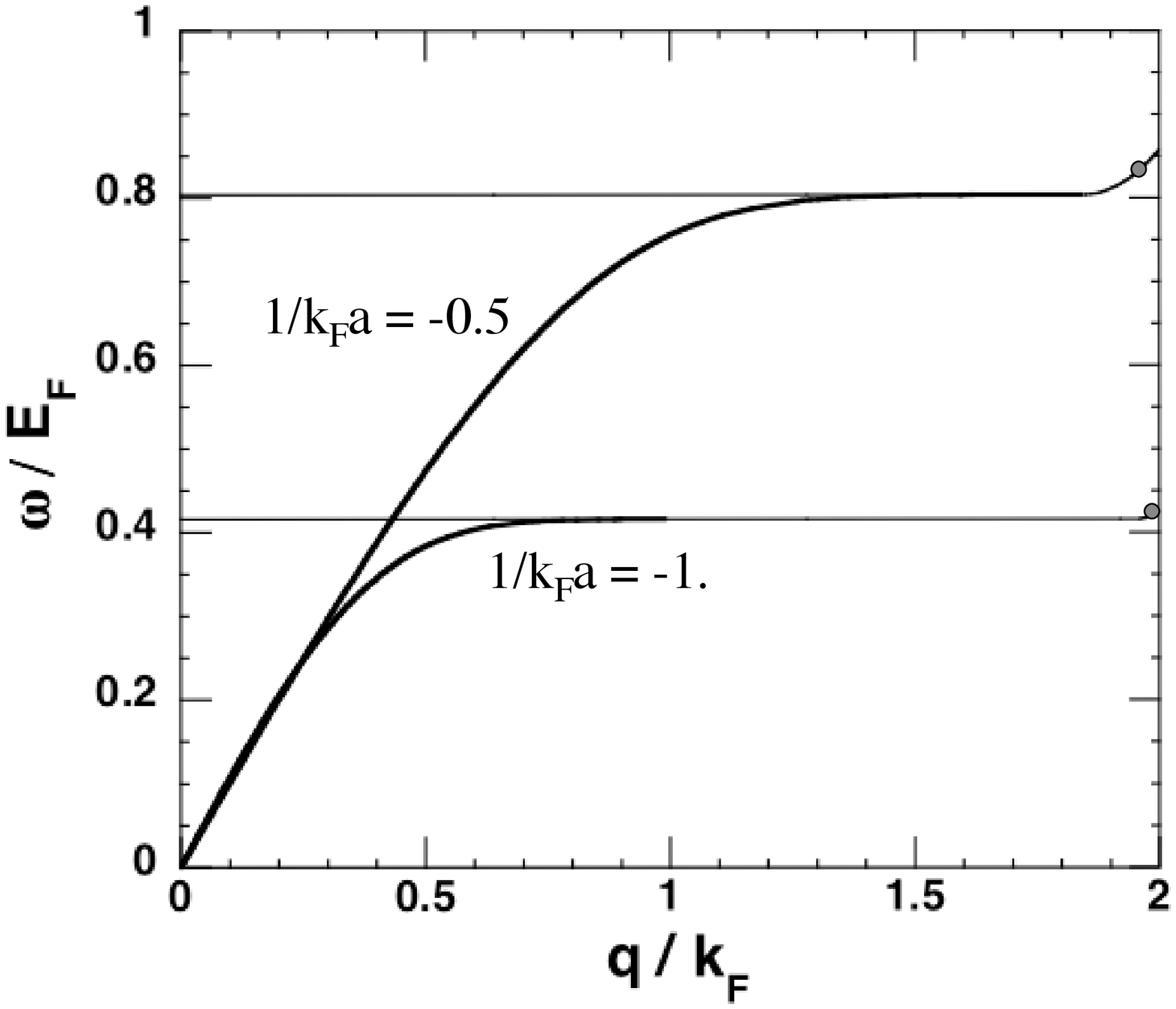}}}}
\vspace{0mm}
\caption{Dispersion relation $\omega /E_F$ of the collective mode as a function of $q/k_F$ for $1/k_Fa= -1. $ (lower thick line) and $-0.5$ (upper thick line). The location of the threshold for pair-breaking is given in each case by the thiner line.}
\label{fig7}
\end{figure}

There is still a strange feature in our result. We have found that the transition toward $\Delta ({\bf r}) \sim\;\cos ({\bf q}.{\bf r})$ is first order in the vicinity of the TCP, while Larkin and Ovchinnikov have a second order phase transition at $T=0$. This implies that, as the temperature is lowered, some change must occur. In this way we have been lead to investigate the problem away from the TCP~\cite{ref:rccm3D}. This is a much more difficult problem since no expansion in powers of $\Delta $ is any longer possible, that is one needs to go to all orders in $\Delta $ to find the actual solution. We have used a method which mixes analytical and numerical ingredients. The final result is the following. When the temperature is lowered, the transition stays first order, that is the qualitative situation is analogous to what we have found in the vicinity of the TCP. However the order parameter changes. At some temperature the best solution switches to a superposition of two cosines along orthogonal directions $\Delta ({\bf r}) \sim  \sum_{i=x,y}\cos ({\bf q}_i.{\bf r})$ and at even lower temperatures it goes to a superposition of three cosines along orthogonal directions $\Delta ({\bf r}) \sim  \sum_{i=x,y,z}\cos ({\bf q}_i.{\bf r})$. This last form is the actual solution at $T=0$. It overtakes the solution found by Larkin and Ovchinnikov. Indeed this solution was beyond their scope since they had restricted their investigation to a second order phase transition.

\section{Collective  mode  in  the  BEC-BCS  crossover}

Let us switch now to our second topic. Since this problem has been in some respect investigated in the literature and that the details of our work, done in collaboration with M.Yu. Kagan and S. Stringari, will be published quite soon~\cite{ref:cks}, I will only stress here the most interesting and new features. We will consider here only the collective mode in a homogeneous superfluid. For trapped gas, the same physics would correspond to the well known collective modes which have been very much investigated experimentally and theoretically. In the low frequency limit this mode reduces to sound propagation with a linear dispersion relation, but we are mostly interested in what happens at higher frequency in the non linear regime. In particular we will consider the merging with the pair-breaking continuum, which in the BEC limit reduces to molecular dissociation continuum. Our considerations will be restricted to $T=0$.

In this BEC limit this mode reduces to the well-known Bogoliubov mode with dispersion relation $\omega ^2 = c_{s}^{2} q^2 + \left(q^2/2M_m \right)^2$. The BCS limit of this mode is less standard, since in superconductors it is pushed up to plasma frequency, because electrons are charged particles, and it becomes physically irrelevant. In superfluid $^3$He, which is a neutral BCS superfluid, the strong hard core repulsion produces a very weak compressibility and a very high sound velocity, which dominates the physics of this mode. So it seems that ultracold fermionic gases are the first systems where the pure Bogoliubov-Anderson (as it is called in this limit) can be observed.

For lack of a much better theory, this collective mode will be investigated within dynamical self-consistent BCS theory. While it is known that this theory gives a proper description both in the BCS and in the BEC limit, it must be considered in between as an interpolation model. However BCS theory is a very coherent theoretical framework, and one may expect that it gives qualitatively correct results. Moreover when we will find a general physical reason supporting our findings, we will gain confidence in their validity.

Let us first consider, in the ($q,\omega $) plane, the domain corresponding to the pair breaking continuum. The threshold $\omega _{th}$ for pair-breaking is given by:
\begin{eqnarray}\label{eqthres}
\omega_{th}=2 {\Delta} \hspace{20mm}\tx{for}\hspace{5mm} \mu  > 0 \hspace{5mm}\tx{and}\hspace{5mm} q \le 2\, \sqrt{2m\mu } \\
\nonumber
\omega_{th}=2 \sqrt{(q^{2}/8m-\mu )^{2}+{\Delta}^{2}} \hspace{28mm}\tx{otherwise}
\end{eqnarray}
and is shown in fig.\ref{fig6} for various values of $1/k_Fa$.
In the BCS limit one recovers the well-known threshold $\omega _{th}=2 \Delta $ when the wavevector is less than $2\,\sqrt{2m\mu }$.
In the BEC limit, one has ${\Delta}/\mu  \rightarrow 0_{-}$ which gives $\omega_{th}=2 |\mu | + q^{2}/4m$. This is just the energy necessary to break a molecule with binding energy $\epsilon _b =  2 |\mu | = 1/ma^2$ into two fermions with total momentum ${\bf q}$, leading to an additional kinetic energy $(1/2) q^{2}/(2m)$, just as for a molecule of mass $2m$.
\begin{figure}
\vspace{-90mm}
\centerline{{\rotatebox{90}{\includegraphics[width=200mm]{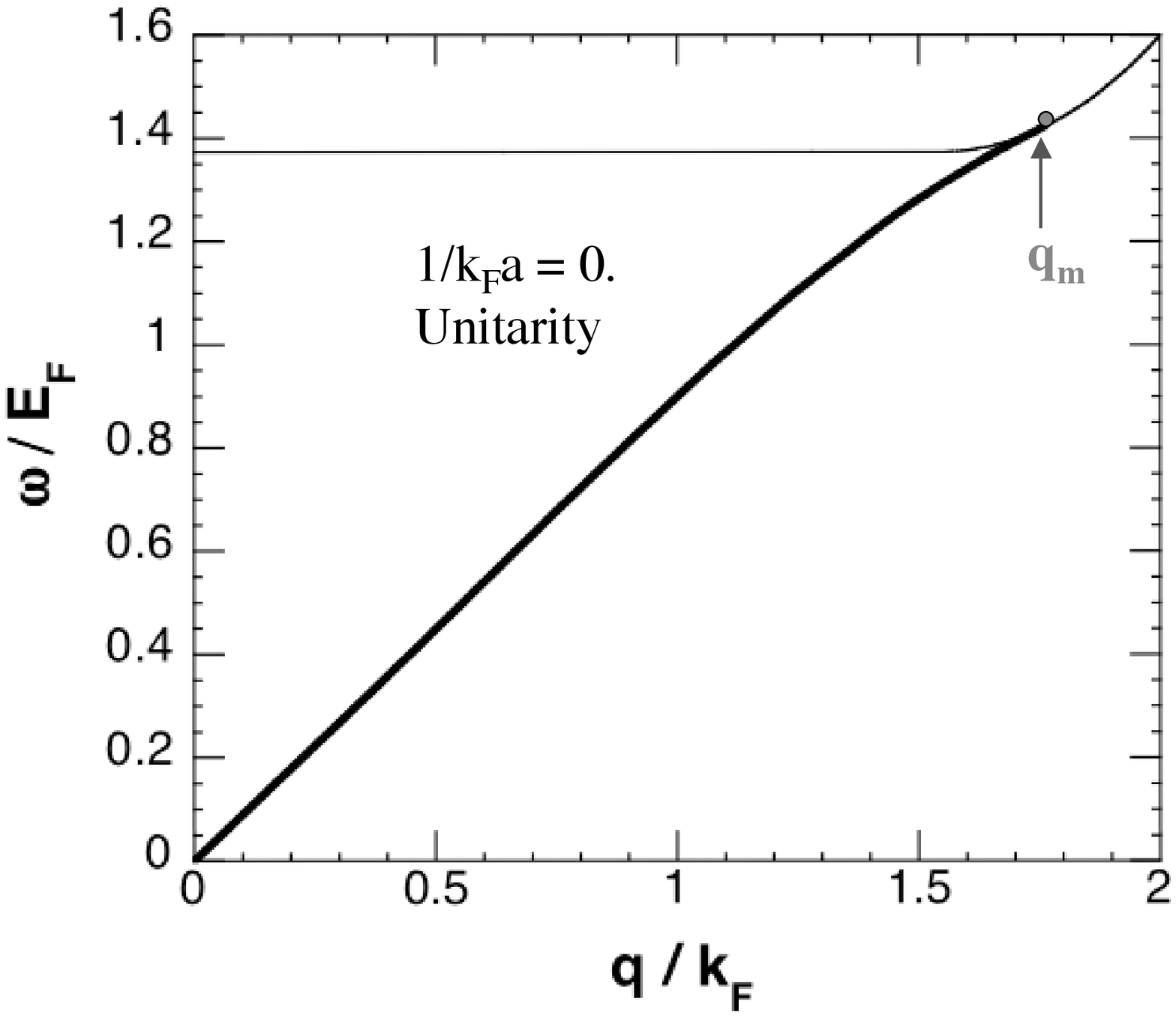}}}}
\vspace{-5mm}
\caption{Dispersion relation $\omega /E_F$ of the collective mode as a function of $q/k_F$ at unitarity (thick line). The location of the threshold for pair-breaking is given by the thiner line. The collective mode merges into the continuum for $q/k_F=1.76$, as it can be seen from Fig.\ref{fig9}.}
\label{fig8}
\end{figure}

We display first in fig.\ref{fig7} the result for the collective mode dispersion relation on the BCS side for two values of $1/k_Fa$. It shows an interesting anticrossing behaviour. This is as if we had a coupled two level system. One of the states is just the collective mode. The other one is the pair breaking continuum. Naturally one can not consider this one as a single state, but the singularity in the BCS density of states at $\omega = 2 \Delta $ means that there is an accumulation of states near this energy, so the physical situation is not so different. Looking at the figure, it seems that the collective mode merges at some point with the threshold $2 \Delta $. Actually this is not correct as shown by detailed analysis. The collective mode approaches the threshold exponentially and merges with the continuum only at higher wavevector, as indicated by the dots near the right side of the figure.
\begin{figure}
\vspace{-60mm}
\centerline{{\rotatebox{90}{\includegraphics[width=200mm]{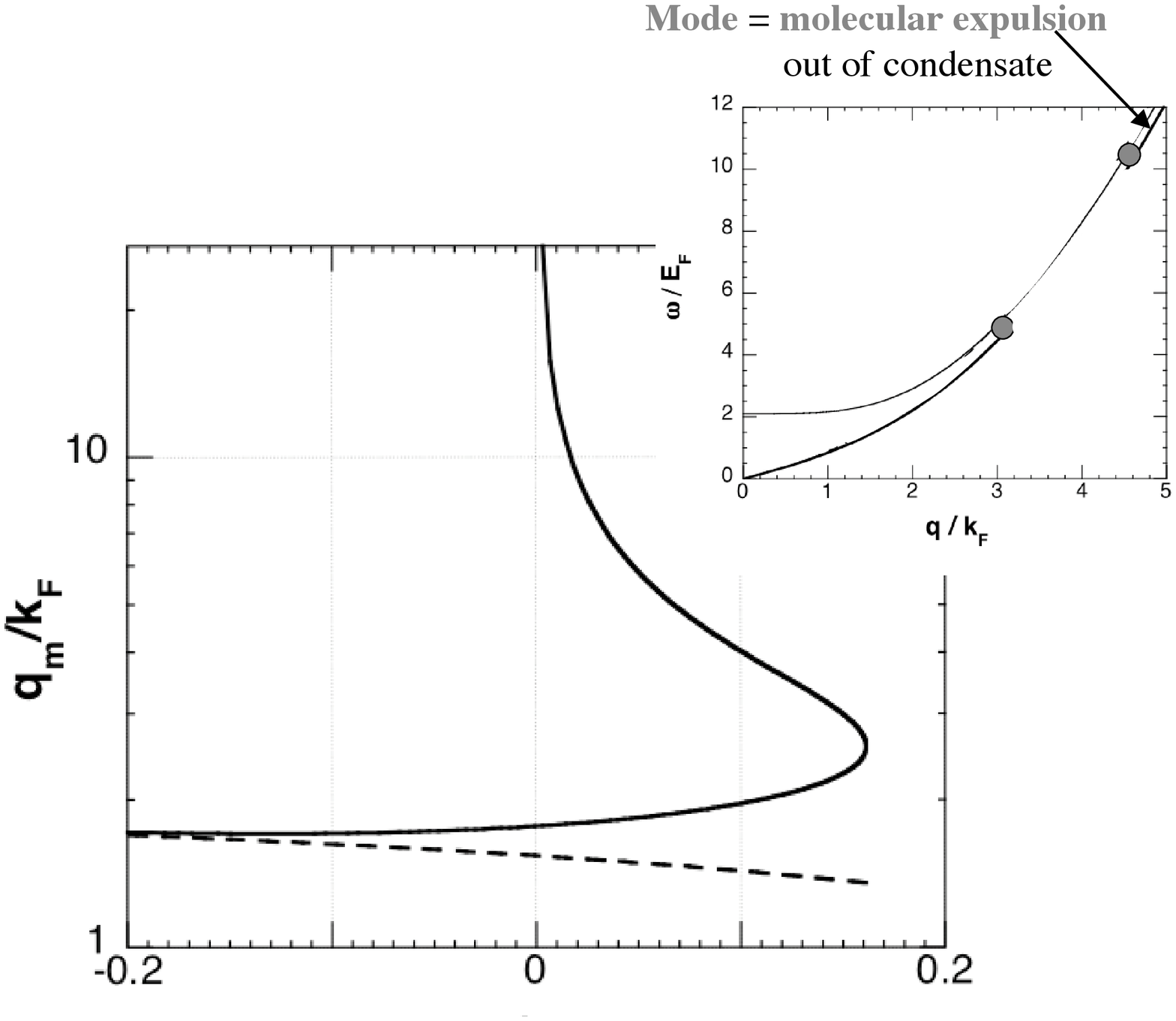}}}}
\vspace{0mm}
\caption{Wavevector $q_m$, in units of $k_F$, at which the collective mode dispersion relation meets the excitation continuum, as a function of $1/k_Fa$ (note that the $q_m$ scale is logarithmic). The dashed line corresponds to $q =2 \sqrt{2m\mu }$. It goes to $q/ k_F=2$ when $a \rightarrow - \infty$. The insert shows schematically the two branches of the collective mode dispersion relation on the $a>0$ side of unitarity.}
\label{fig9}
\end{figure}

We consider next the dispersion relation at unitarity, which is found in fig.\ref{fig8}. Qualitatively the result is rather similar to the one found on the BCS side. However the interesting point here is the very wide range of quasi linear behaviour for this dispersion relation. This may explain why hydrodynamics works so well in traps around unitarity at the fairly high frequencies produced in radial oscillations, since $T=0$ hydrodynamics is valid as long as the dispersion relation is linear.
\begin{figure}
\vspace{-60mm}
\centerline{{\rotatebox{90}{\includegraphics[width=200mm]{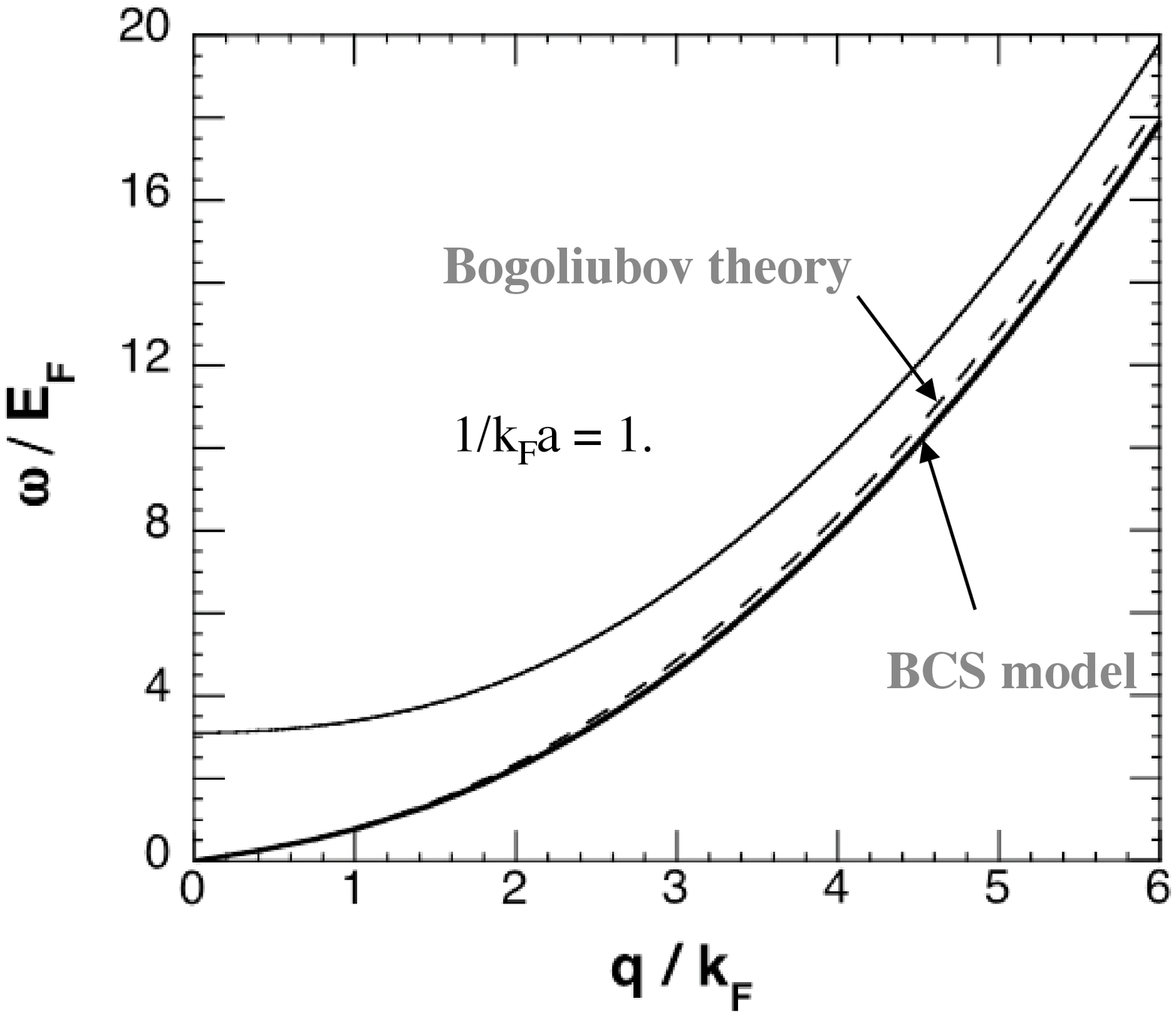}}}}
\vspace{-5mm}
\caption{Dispersion relation $\omega /E_F$ of the collective mode as a function of $q/k_F$ for $1/k_Fa  = 1.$ (thick line). The threshold for pair-breaking is given by the thiner line. The dashed line indicates the result obtained from the Bogoliubov formula.}
\label{fig10}
\end{figure}

We turn now to the specific location of the wavevector $q_m$ where the collective mode merges with the pair breaking continuum. A natural expectation would be that $q_m \rightarrow \infty$ as one approaches unitarity. This would allow to recover on the BEC side a collective mode which never meets the continuum, as it is the case for the Bogoliubov mode. The fact that things are more complicated is already plain from fig.\ref{fig8} where one sees that, at unitarity, merging occurs still at a finite value of $q_m$. The whole picture is given in fig.\ref{fig9}. Surprisingly a higher frequency branch of the collective mode appears at unitarity, which merges with the continuum at lower frequency, as it is indicated schematically in the insert of  fig.\ref{fig9}. Hence on the BEC side, that is for $a>0$, there are two merging points $q_m$. The evolution of their position with $1/k_Fa$ is indicated in the main figure. We see that, when $1/k_Fa$ increases, they come closer until they become equal. At this stage the collective mode dispersion relation only touches the continuum. For larger $1/k_Fa$ one recovers a single dispersion relation located below the continuum, that is the same qualitative situation as the standard Bogoliubov mode. The interpretation of this strange behaviour is the following. Physically the Bogoliubov mode at high frequency corresponds to kick a molecule out of the condensate. However it has been pointed out~\cite{ref:clk} that, while molecular formation goes unhindered at high momentum, it becomes restricted by the effect on the molecular state of Pauli exclusion by the Fermi sea made of the other fermions. Hence, near unitarity, molecules can not form at lower momentum and the collective mode can not exist. We note that, since these arguments do not rely specifically on BCS theory, we expect this behaviour to be qualitatively present in the exact theory.

Finally in the last figure we compare in the BEC regime our result for the collective mode dispersion relation with the standard Bogoliubov. We see that, while they agree perfectly well at low frequency (which is quite natural since the sound velocity is the same in the two results), the Bogoliubov result is slightly higher than the collective mode we find. This can be understood qualitatively because, at higher momentum $q \sim 1/a$, the collective mode will feel the internal structure of the molecules whose size is of order $a$. Hence the fact that our molecules are composite bosons will appear. Since Bogoliubov theory deals only with elementary bosons, it is quite natural that it does not lead to exactly the same result in this high $q$ regime. Again since this physical argument does not make use specifically of BCS theory, we expect this result to hold also in the exact theory of this regime. 

\acknowledgments
We are grateful to C. Mora for providing some of the figures used in this paper.

\end{document}